\documentclass[prd,aps,onecolumn,preprintnumbers,nofootinbib,showkeys,showpacs,]{revtex4}

\usepackage[T2A]{fontenc}
\usepackage{amsmath,amssymb}
\usepackage{eucal}
\usepackage{color}
\usepackage[dviwindo]{graphicx}
\newcommand{\bs}{\begin{subequations}}
\newcommand{\es}{\end{subequations}}
\numberwithin{equation}{section}
\newcommand{\ben}{\begin{eqnarray}}
\newcommand{\een}{\end{eqnarray}}
\newcommand{\la}{\label}

\begin{document}

\title{A realistic model of a neutron star in minimal dilatonic gravity}

\author{Plamen P. Fiziev
\footnote{fizev@phys.uni-sofia.bg\,\,\,and\,\,\,
fizev@theor.jinr.ru}}
\affiliation{BLTF, JINR, Dubna, 141980 Moscow Region, Russia}

 \begin{abstract}

 We present a derivation of the basic equations and boundary conditions
 for relativistic static spherically symmetric  stars (SSSS)
 in the  model of minimal dilatonic  gravity  (MDG) which offers an alternative and simultaneous
 description of the effects of dark matter (DM) and dark energy (DE) using one dilaton field $\Phi$.
 The numerical results for a realistic equation of state (EOS) MPA1 of neutron matter are presented for the first time.
 The three very different scales, the Compton length of the scalar field $\lambda_\Phi$,
 the star's radius $r^*$, and the finite radius of the MDG Universe $r_{U}$,
 are a source of numerical difficulties.
 Owing to the introduction of a new dark scalar field $\varphi=\ln(1+\ln\Phi)$,
 we have been able to study numerically an unprecedentedly large interval of $\lambda_\Phi$
 and have discovered the existence of $\lambda_\Phi^{crit}\approx 2.1$\ km for a neutron star with MPA1 EOS.
 This is related to the bifurcation of the physical domain in the phase space of the system.
 Some novel physical consequences are discussed.


\end{abstract}
\sloppy
\maketitle
\section{Introduction}\label{Intro}

One of the most important lessons from the spectacular development of cosmology
in the last fifteen years is the clear understanding that Einstein's General Relativity (GR),
as a model of gravity, and the Standard Particle Model (SPM), as a model of matter,
are not enough to explain all the observed phenomena in Nature.

There exist three possible ways for further development:
a) To add some new content to the Universe, such as DM and DE;
b) To change the theory of gravity\footnote{The two simplest modifications are the $f(R)$ model, see for example
\cite{Buhdahl72,Starobinsky80,Muiller88,Nojiri03,Nojiri04,Carroll04,Abdalla05,Capozzielo06a,Capozziello06b,
Faraoni06,Starobinsky07,Amendola07,Appleby07,Hu07,Li07,Bamba08,Amendola08,Nojiri08,Cognola08,Tsujikawa08,Nojiri08c,
Starobinsky10,Felice10,Sotiriou10,Tsujikawa10,Motohashi10,Capozziello11,Nojiri11,Capozziello11a,Rubakov11,Clifton12,Bamba12,Gannouji12,Starobinsky15} and MDG \cite{OHanlon72,Fiziev00a,EFarese01,Fiziev02,Fiziev03,Fujii03,Fiziev13}.
These are similar, but not identical.
Much more sophisticated models of modified gravity are also under investigation at present, see for example the references in \cite{Capozziello11,Capozziello11a,Berti2015}.
All of them have a large number of additional parameters without clear physical justification
and will be not considered here.};
c) The current observational data do
not exclude a combination of a) and b).

While the need for DM and DE has already been firmly established \cite{Planck_XIV_15},
their nature and their small-scale distribution are still largely unknown.
The only settled part  concerning  DM is its gravitational interaction.
We have no evidence that DM has any other interaction but gravitational.
We also have no idea what is the nature of DE. Within the framework of MDG \cite{Fiziev03}
DE can be thought of as responsible for two quite different physical processes with similar manifestations:
the initial inflationary expansion and the present accelerating expansion of the Universe.

The main general problem for the construction of $f(R)$ theories still remains the absence
of physical intuition when we are trying to specify the function $f(R)$ \cite{Buhdahl72}.
Nowadays, a series of additional requirements have been formulated
aiming at cosmological applications \cite{Starobinsky10,Starobinsky15}.
However, in the literature one can find dozens of such functions.
Several of them, for example \cite{Starobinsky80,Starobinsky07,Appleby07,Hu07},
are thought of as valuable, bearing in mind cosmological applications.

The situation in star physics is similar.
The development so far, which has already lasted several decades,  has not solved the problem of finding the real EOS
of dense matter. At present, one can find several dozens of them dubbed realistic in the literature,
see for example the very recent review \cite{Burgio2015}.

There have also been a series of attempts to use  $f(R)$ models of gravity adapted to star physics\footnote{See \cite{Kainulainen07a,Kainulainen07b,Multamaki07,Nojiri07,Multamaki08,Frolov08,Kobayashi08,Henttunen08,Kainulainen08,Nojiri09,Babichev09,Sotiriou10,Babichev10,Cooney10,Capozziello11,Capozziello11b,Nojiri11,Arapoglu11,Santos12,Ang12,Deliduman12,Alavirad13,Astashenok13,Ganguly13,Orellana13,Astashenok14,Astashenok15} and the references therein.}.
For some constraints on $f(R)$ for star models and specific numerical solutions, see \cite{Babichev09,Babichev10}.
No convincing final result has been reached. In the recent review \cite{Berti2015} one can find a
description of the present state of affairs:
While it is hard to construct Neutron Star (NS) equilibrium configurations in $f(R)$ gravity
from a numerical point of view, there is no fundamental obstacle to their existence.
NS configurations with realistic values of the physical parameters have never
been constructed in viable $f(R)$ models.

The main goal of the present paper is to create a clear physical, analytical and numerical basis
for the application of one of the simplest modification of GR, namely MDG,
and to present for the first time a realistic model of NS in it.

We hope that this consideration may help future developments of MDG models
of the (almost) spherical objects at very different scales:
laboratory scales, compact star scales, and at the scales of planets, white dwarfs,
standard stars, star clusters, dwarf sphericals, galaxies,
and clusters of galaxies.
The use of the available information for similar physical phenomena at all reachable scales
will give a much more definite justification of the model.
A simultaneous and coherent adequate description of these phenomena
is a promising  way to overcome the existing problems
that are related to the absence of sufficient experimental and observational information
about both matter and gravity.
Such an ambitious program needs a well-developed theoretical, analytical and numerical basis.
The best way to start is to probe well-known and simple realistic examples of SSSS, like MDG-NS
\cite{Fiziev14a,Fiziev14b,Fiziev15a,Fiziev15b}.

The MDG model was proposed and studied in \cite{OHanlon72,Fiziev00a,Fiziev02,Fiziev13}.
It describes a proper simple modification of GR
based on the action ${\cal A}= {\cal A}_{g,\Phi}+{\cal A}_{matter}$.
The action of the gravi-dilaton sector is
\ben
{\cal A}_{g,\Phi}={\frac c {2\kappa}}\int d^4 x\sqrt{|g|}
 \bigl( \Phi R - 2 \Lambda U(\Phi) \bigr).
\la{A_MDG}
\een
We take the Einstein constant $\kappa=8\pi G_{N}/c^2 \approx 1.8663\times 10^{-27}\,cm\, g^{-1} $, Newton's constant $G_N\approx 6.6738 \times10^{-8}\,cm^3 g^{-1} s^{-2}$,
and the cosmological constant $\Lambda \approx 1.0876\times10^{-56}\, cm^{-2}$. The dilaton field is $\Phi\in (0,\infty)$.
In general, this model is only locally equivalent to the $f(R)$ model \cite{Fiziev13}, and has a clear physical meaning:
\begin{itemize}
  \item The scalar field $\Phi$ is introduced to take into account a variable gravitational factor $G(\Phi)=G_N/\Phi=G_N g(\Phi)$ instead
of the gravitational constant $G_N$ and does not enter into ${\cal A}_{matter}$, having no interaction with SPM matter.
  \item The cosmological potential $U(\Phi)$ is introduced so as to have a variable cosmological factor $\Lambda U(\Phi)$
instead of the cosmological constant $\Lambda$.
\end{itemize}
In GR with cosmological constant $\Lambda$, we have $\Phi\equiv 1$, $g(\Phi)\equiv 1$, and $U(1)\equiv 1$.
Due to its specific physical meaning, the field $\Phi$ has quite unusual mathematical and physical properties
and does not enter into the standard action ${\cal A}_{matter}$.

For astrophysical reasons, the cosmological potential $U(\Phi)$ must be a positive single valued function of $\Phi \in (0,\infty)$.
In  \cite{Fiziev13}, there was introduced the class of {\em withholding} potentials,
in order to confine dynamically the values of the dilaton $\Phi$ in the physical domain,
excluding antigravity, ghosts and tachyons.
It is hard to formulate an analogous general requirement for $f(R)$.
Thus, we immediately see the main advantage of the MDG model, even when it is formally equivalent to some specific $f(R)$ theory:
We have a great deal of physical experience working with potentials like the cosmological one,
both at the level of classical and quantum mechanics, or classical and quantum field theory.

In units where $G_N=c=1$, the field equations of MDG can be written in the form\footnote{Here ${T}_\alpha^\beta$ is the standard conserved energy--momentum tensor of the matter,
$\hat{X}_\alpha^\beta={X}_\alpha^\beta - {\frac 1 4}X\delta_\alpha^\beta$
is the traceless part of any tensor ${X}_\alpha^\beta$ in four dimensions,
$X=X_\alpha^\alpha$ is its trace,  and the comma denotes differentiation with respect to $\Phi$.}:
\ben
\Phi \hat{R}_\alpha^\beta +\widehat{\nabla_\alpha\nabla^\beta}\Phi+ 8\pi\hat{T}_\alpha^\beta=0, \quad
\Box\Phi+ \Lambda V_{\!{}_{,\Phi}}(\Phi) = {\frac {8\pi} 3} T.
\label{DGE}
\een

The relation $V_{\!{}_{,\Phi}}(\Phi)={\frac 2 3}\Big( \Phi\, U_{\!{}_{,\Phi}}(\Phi) -2U(\Phi) \Big)$
defines the dilatonic potential $V(\Phi)$
(For the conventions used, see \cite{Fiziev13}).

The main physical problem with all modifications of GR with one (or more) additional scalar field $\Phi$ is
the value of its mass $m_{\Phi}$. In the simplest cases of modified gravity, this is the only new parameter.
This problem appeared for the first time as early as in \cite{Fujii71} and is still remains unsolved.

In Starobinsky (1980) presents an $f(R)$ theory with one additional parameter\cite{Starobinsky80} $f(R)=R+R^2/\,6m_{\Phi}^2$.
Despite the fact that this model is still the most successful model of initial inflation \cite{Planck_XX_15}, its potential $V(\Phi)\sim (\Phi-1)^2$ allows antigravity and makes the model unacceptable in different physical situations.
As longas one is working in a small enough vicinity of the vacuum state $\Phi_{vac}=1$, this shortcoming may be ignored at the classical level.
In general, it will be  not ignorable at the quantum level.
Comparing the scalaron mass with the cosmological data about the initial inflation, Starobinsky
was able to find the value $m_{\Phi}\sim 3\times 10^{-6}\,M_{Plank}$ \cite{Starobinsky07}.
This gives an extremely small Compton length $\lambda_{\Phi}\sim 10^{-27}$\ cm of the scalar field.
Such a model is indistinguishable from GR at the scales which we discussed above
because of the Yukawa character of the corrections ($\sim exp(-r/\lambda_{\Phi})$) to Newtonian gravity
\cite{Fujii71,OHanlon72,Fiziev00a, Fujii03,Murata15}.

On the other hand, in the linear approximation, MDG reproduces the Yukawa tails to Newton's law. The comparison with
the laboratory and solar system experiments known in 1999 led to the estimate $m_{\Phi} \geq 10^{-3} eV/c^2$, which corresponds to
$\lambda_{\Phi}\sim 10^{-2}$\ cm \cite{Fiziev00a}.
Later on, similar results were reproduced and refined many times in the framework of $f(R)$ theories,
see for example the recent papers \cite{Starobinsky10,Starobinsky15}.
According to the latest review of the experimental data \cite{Murata15},
the Compton length of the scalar field is $\lambda_{\Phi}< 2.3 \times 10^{-3}$\ cm.

The simplest withholding dilaton potential in MDG can be written in the form\footnote{The potential
\eqref{V} has an unique minimum (the de Sitter vacuum) at $\Phi_{vac}=1$ and is
the only withholding potential for which the corresponding Newtonian-like equation in a flat space time
is solved by elliptic functions and the corresponding Schr\"odinger-like equation
is solved by Heun's functions.
All other withholding potentials in these two cases will require the use of hyper-elliptic functions
or of the not well-studied  Fuchsian functions with more than four singularities.} \cite{Fiziev02,Fiziev03,Fiziev13}:
\ben
V(\Phi)={\frac 1 {2 \mathfrak{p}^2}}\left(\Phi+1/\Phi-2\right),
\la{V}
\een
and used in NS physics in \cite{Fiziev14a,Fiziev14b,Fiziev15a,Fiziev15b}.
Here $\mathfrak{p}=\lambda_{\Phi}\sqrt{\Lambda}$ is
the dimensionless Compton length of the scalar field in cosmological units.
According to the above estimates, this extremely small quantity lies
in the physical interval $\mathfrak{p} \in (1\times 10^{-55},2.4\times 10^{-31})$.
Taking into account that the largest value was obtained only in the linear approximation,
one can not be sure that much larger values of $\mathfrak{p}$ are excluded by observations.
Actually, to explain the current accelerating expansion of the Universe in the framework of the quintessence models,
or the modified gravity models equivalent to them, one needs a very small mass of the scalar field
$m_\Phi\sim 2 \times 10^{-33}$\ eV \cite{Wetterich15}. This gives
$\lambda_\Phi\sim 10^{25}$\ cm and  $\mathfrak{p}\sim 10^{-3}$.
The last value is still admissible in MDG \cite{Fiziev02,Fiziev03}.
Hence, to check the MDG model at different physical scales, we need special
techniques to deal with the interval $\mathfrak{p} \in (10^{-55},10^{-3})$.
This is a very challenging task\footnote{One way to get around this problem is to use
some sophisticated additional mechanisms like the chameleon, K-mouflage, or Vainshtein mechanisms,
see for example \cite{Khoury04,Brax15} and the references therein.
Then the mass $m_\Phi$ depends on the environment in a somewhat ad hoc and artificial way.}.

In MDG we can overcome the physical problem of the existence of different masses $m_\Phi$
by introducing the potential $V(\Phi)$ with many minima \cite{Fiziev13}.
Then, around each of these minima, a Taylor series expansion will produce different effective masses $m_\Phi$ of the scalar field.
If so, the first step will be to study problems with simple potentials \eqref{V} for different values of the parameter
$\mathfrak{p}$, and then to try potentials with many minima.
One can hope that different effective values of $m_\Phi$ will correspond
to the above problems with different physical scales.

In the present paper we study the problem at the scale of realistic NS.
One of the goals is to recover the reasons for the numerical difficulties
and to develop new methods to surmount them.

We solve some of the physical problems by introducing the notions of cosmological energy density and pressure,
and dilatonic energy density and pressure, see Eqs. \eqref{epL} and \eqref{epP},
as well as novel equations of state for them: Cosmological EOS (CEOS) and Dilatonic EOS (DEOS),
which have to be used in MDG together with the familiar Matter EOS (MEOS)
\cite{Fiziev14a,Fiziev14b,Fiziev15a,Fiziev15b}.
An essential role is played by the finiteness of the dilaton pressure at the center of a star,
see Eqs. \eqref{center}.
Here we show for the first time in detail how these notions and relations arise from MDG (See Appendix \ref{basicEqsa}.).

A star is a matter excitation above the proper vacuum state of the theory.
Since in MDG the physical vacuum is the de Sitter vacuum \cite{Fiziev02,Fiziev13},
we define the surrounding mass $m(r)$ of the star using the relations $-1/g_{rr}=1-2m(r)/r-{\tfrac 1 3}\Lambda r^2$
\cite{Fiziev14a,Fiziev14b,Fiziev15a,Fiziev15b}.
Note that, in those articles on stars in $f(R)$ models
known to the author, the mass $m(r)$ is often defined using the incorrect assumption that the vacuum state
corresponds to a flat spacetime asymptotic, i.e., without the $\Lambda$ term.

Outside the star, the dilaton creates a ``dilaton sphere,'' or, for short,  a ``disphere''
\cite{Fiziev14a,Fiziev14b,Fiziev15a,Fiziev15b}.
In the non spherically-symmetric case, one can speak of a ``dark domain'' outside the matter.
The disphere exponentially dilutes up to the radius of the Universe $r_U$, defined by the cosmological horizon of the MDG model
where the de Sitter vacuum is reached.
Here, we calculate, for the first time, the mass of the disphere of NS with a realistic MEOS,
and the total mass $m_{total}$ of the whole complex: NS plus disphere.
Some similar results and a not very precise use of the terminology
of articles \cite{Fiziev14a,Fiziev14b,Fiziev15a,Fiziev15b}
can be found in \cite{Astashenok15} devoted to quark stars with MIT-bag-model MEOS.

\section{Basic equations and boundary conditions for SSSS in MDG}

The spacetime interval for SSSS is
$ds^2=e^{\nu(r)}dt^2-e^{\lambda(r)}dr^2 - r^2 d\Omega^2 $,
where $r$ is the luminosity distance to the center of symmetry
and $d\Omega^2$ describes the space-interval on the unit sphere\footnote{The luminosity radius
$r$ is an invariant defined by the relation $A=4\pi r^2$;
$A$ is the area of a sphere around the center of symmetry.}.
After some algebra (See Appendix \ref{basicEqsa}) one obtains  the following basic results
for SSSS:

In the inner domain   $r\in [0,r^*]$, the SSSS structure is determined by
the fourth order generalization of the TOV equations:
\begin{subequations}\label{DE:ab}
\ben
{\frac {dm}{dr}}&=&4\pi r^2\epsilon_{eff}/\Phi,\quad
{\frac {dp}{dr}}=- {\frac {p+\epsilon}{r}}\,{\frac{m+4\pi r^3 p_{eff}/\Phi}{\Delta-2\pi r^3 p_{{}_\Phi}/\Phi} }, \label{DE:a}\\
{\frac {d\Phi}{dr}}&=&-4\pi r^2 p_{{}_\Phi}/\Delta,\quad
{\frac {dp_{{}_\Phi}}{dr}}=- {\frac{ p_{{}_\Phi}}{r\Delta}}\left(3r -7 m-{\frac 2 3}\Lambda r^3+4\pi r^3\epsilon_{eff}/\Phi\right)
-{\frac{2}{r}}\epsilon_{{}_\Phi}. \label{DE:b}
\een
\end{subequations}
The four unknown functions are $m(r)$, $\Phi(r)$, $p_{{}_\Phi}(r)$, and $p(r)$. In Eqs. \eqref{DE:ab}
$\Delta(r)=r-2 m(r)-{\frac 1 3}\Lambda r^3$,
$\epsilon_{eff}=\epsilon+\epsilon_{{}_\Lambda}+\epsilon_{{}_\Phi}$,
$p_{eff}=p+p_{{}_\Lambda}+p_{{}_\Phi}$.
In addition to the standard MEOS $\epsilon = \epsilon(p)$, we obtain two novel equation:
CEOS: $\epsilon_{{}_\Lambda}=- p_{{}_\Lambda}-{\frac \Lambda {12\pi}}\Phi$ and
DEOS: $\epsilon_{{}_\Phi}=p-{\frac 1 3}\epsilon +
{\frac \Lambda {8\pi}}V_{\!{}_{,\Phi}}(\Phi)+{\frac {p_{{}_\Phi}} 2}\,{\frac{m+4\pi r^3 p_{eff}/\Phi}{\Delta-2\pi r^3 p_{{}_\Phi}/\Phi}}$.

The cosmological energy density and the cosmological pressure are defined as follows:
\ben
\epsilon_{{}_\Lambda} ={\frac \Lambda {8\pi}} \Big(U(\Phi)-\Phi\Big),\quad
p_{{}_\Lambda} =-{\frac \Lambda {8\pi}} \Big(U(\Phi)-{\frac 1 3}\Phi\Big).
\la{epL}
\een

The dilatonic energy density and the dilatonic pressure measure the changes of the gravitational factor.
By definition:\footnote{Here we use the true geometrical radial distance
$ l(r) = \int_0^r {\frac {dr} { \sqrt{1-{\frac {2m(r)} r}-{\frac \Lambda 3}r^2 }}}$.
It monotonically increases from 0 to some finite value $l_U> r_U$ (the size of the Universe), since ${\frac{dl}{dr}\geq 1}$;
$l(r)\sim r$, when $r\to 0$; and ${\frac{dl}{dr}\to \infty}$, when $r\to r_U$. }
\ben
\epsilon_{{}_\Phi}={\frac 1 {8\pi}}{\frac 1 A} {\frac {d}{dl}}\left(\!A\,{\frac {d\Phi}{dl}}\right)=
{\frac 1 {8\pi r^2}}\left(\Delta/r\right)^{1/2}\left(r^2 \left(\Delta/r\right)^{1/2}\Phi^\prime\right)^\prime,
\quad
p_{{}_\Phi}=-{\frac {\sqrt{-g_{rr}}} {4\pi r}}\,{\frac {d\Phi}{dl}}=-{\frac \Delta {4\pi r^2}}\Phi^\prime.
\la{epP}
\een

Placing the physical SSSS-center at $r_c=0$,
we obtain the boundary conditions (See Appendix \ref{basicEqsb}.):

\ben
m(0)=m_c=0,\,\,\,\,\,\Phi(0)=\Phi_c,\,\,\,\,\,p(0)=p_c,\,\,\,\,\,
p_{{}_\Phi}(0)=p_{\Phi c}={\frac 2 3}\left({\frac {\epsilon(p_c)} 3}- p_c\right)-{\frac{\Lambda}{12\pi}}V_{\!{}_{,\Phi}}(\Phi_c).
\la{center}
\een
Requiring $m_c=0$ ensures the finiteness of the pressure $p_c$ simultaneously for the Newton-, GR- and MDG-SSSS.
The condition on $p_{\Phi c}\,(=-{\frac 2 3}\epsilon_{\Phi c})$  ensures its finiteness,
being a specific relation for the values of the MDG-center: $F_\Phi(p_{\Phi c},p_c,\Phi_c)=0$.

The SSSS-edge  is defined by the condition $p^*=p(r^*;p_c,\Phi_c)=0$. Then
\ben
m^*=m(r^*;p_c,\Phi_c),\,\,\,\Phi^*=\Phi(r^*;p_c,\Phi_c),\,\,\,
p_{{}_\Phi}^*=p_{{}_\Phi}(r^*;p_c,\Phi_c).
\la{edge}
\een
The luminosity radius of a compact star with physically realistic MEOS may vary: $r^*\approx 8\div 14$ km.

To obtain a complete description of the spacetime geometry inside SSSS, one
must add Eq. \eqref{dnu}, which splits out from Eqs. \eqref{DE:ab},
as well as the corresponding boundary conditions for $\nu(r)$.

\begin{figure}[!ht]
\begin{minipage}{9.cm}
\vskip -.truecm
\hskip -.truecm
\includegraphics[width=10.truecm]{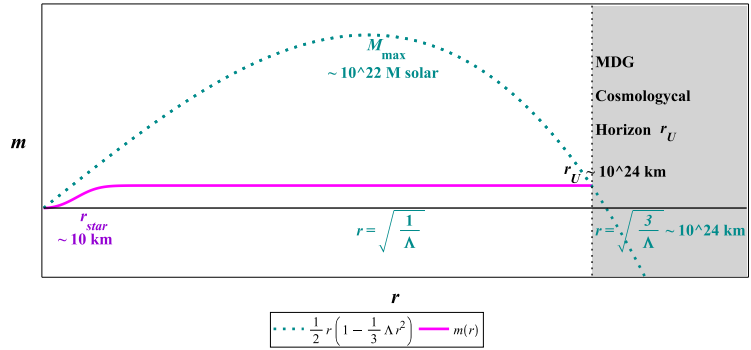}
\vskip .truecm
\caption{\small A schematic picture of MDG Universe with one SSSS}
\label{Fig1}
\end{minipage}
\end{figure}

Outside the star, $p\equiv 0$ and $\epsilon\equiv 0$, and we have a disphere \cite{Fiziev14a,Fiziev14b,Fiziev15a,Fiziev15b}.
Its structure is described by the shortened system \eqref{DE:ab}, where  the second of Eqs. \eqref{DE:a} is omitted.
For the exterior domain $r\in [r^*,r_{{}_{\!U}}]$, we use Eqs. \eqref{edge} as left boundary conditions.
The right boundary is defined by the MDG cosmological horizon $r_{{}_{\!U}}$:  $\Delta(r_{{}_{\!U}}; p_c,\Phi_c)=0$,
where the de Sitter vacuum is reached:
$\Phi(r_{{}_{\!U}}; p_c,\Phi_c)=1$ and $g_{tt}=0$, $g_{rr}=\infty$, $g_{tt}\,g_{rr}=1$.
As a result, we obtain a new relation for the values of the MDG SSSS center, $F_\Lambda(p_c,\Phi_c)=0$
which depends also on the Compton length $\lambda_\Phi$.
Besides, the 4d and 3d scalar curvatures are ${}^{(4)}\!R=4\Lambda-1/r_U^2$ and ${}^{(3)}\!R=4\Lambda$.
A schematic picture of MDG Universe with only one SSSS is shown in Fig.~\ref{Fig1}.

The two MEOS dependencies
$F_\Phi(p_{\Phi c},p_c,\Phi_c)=0,\,\,\,F_\Lambda(p_c,\Phi_c)=0$
show that for a given MEOS in MDG, as well as in Newtonian gravity and GR, we have a one-parameter family of SSSSs.

Now it becomes clear that a basic difficulty in the numerical investigation of NS in MDG is the presence of three very different
scales: $\lambda_\Phi \in (10^{-27}\, cm , 10^{25}\, cm)$, $r^* \sim 10^6 \,cm$, and $r_{U} \sim 10^{28}\,cm$.
The justification of the values of $\lambda_\Phi$ for different astrophysical objects becomes the most important
physical problem for modified gravity.

In the MDG Universe, the maximum mass of any matter object is $M_{max}={ \frac {8\pi} {\kappa\sqrt{\Lambda}} }\sim 10^{22} M\odot$.
This value is physically safe as it is about six orders of magnitude greater than the mass of the most massive objects in the Universe \cite{Holz12}.

\section{A Model of a Neutron Star with MEOS MPA1}
We use the well-known realistic MEOS  MPA1 \cite{MPA1a,MPA1b,MPA1c}.
Its analytic version \cite{MPA1c} allows a treatment of NSs with central densities up to $10^{15}$\ g/cm$^3$.
The matter density decreases to $\rho_{{}_{Fe^{56}}}\approx 6.5$\ g/cm$^3$ on the surface of the NS.

To have successful computations, we were forced to implement high-precision computer arithmetic with 64 digits
and to replace the dilatonic field $\Phi$ with a novel field variable
\ben
\varphi=\ln(1+\ln\Phi) \Leftrightarrow \Phi=\exp(\exp(\varphi)-1).
\la{DS}
\een
We call the field $\varphi$ {\em the dark scalar}.
Note that the mass of the dark scalar precisely equals the mass of the dilaton $m_\Phi$.

The double logarithmic substitution \eqref{DS} stretches the physical domain of the scalar field
and makes possible numerical calculations in the maximal interval for the parameter $\mathfrak{p}$ allowed by the given MEOS.
Inside the star we use standard logarithmic variables $\xi=\log(\rho)$ and $\zeta=\log(p)$.
Outside the star we also use a proper logarithmic variable instead of the luminosity radius $r$.
The numerical results were obtained by an appropriate version of the shooting method.

\begin{figure}[!ht]
\begin{minipage}{18.cm}
\vskip -.truecm
\hskip -.truecm
\includegraphics[width=6.truecm]{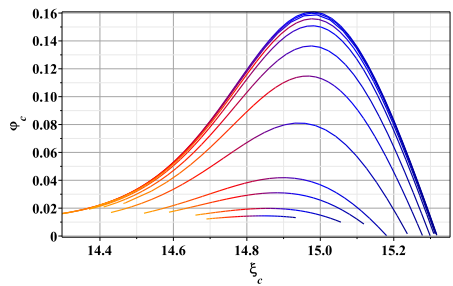}
\vskip -.2truecm
\caption{\small The specific MDG curves $F_\Lambda(\varphi_c,\xi_c;\lambda_\Phi)=0$ for NS with MPA1 MEOS and different fixed
$\lambda_\Phi \in (2.4$\ km$, 10^4$\ km). The curves, from top to bottom, correspond to decreasing values of $\lambda_\Phi$.}
\label{Fig2}
\end{minipage}
\end{figure}

\begin{figure}[!ht]
\begin{minipage}{12.cm}
\vskip -.3truecm
\hskip .4truecm
\includegraphics[width=6.truecm]{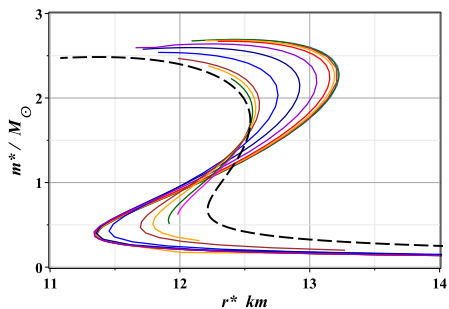}
\vskip .truecm
\caption{\small The MDG mass--radius relation for a NS with MPA1 MEOS for different fixed $\lambda_\Phi \in (2.4, 10^4)$\ km.
The black dashed curve presents the GR mass--radius relation}
\label{Fig3}
\end{minipage}
\end{figure}

\begin{figure}[!ht]
\begin{minipage}{12.cm}
\vskip -.truecm
\hskip .5truecm
\includegraphics[width=6.truecm]{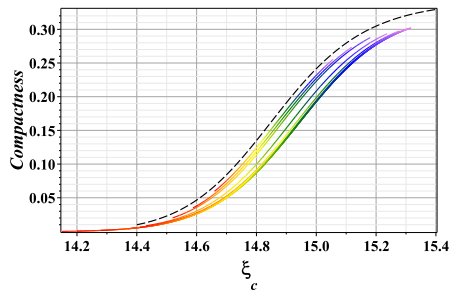}
\vskip .truecm
\caption{\small The compactness of an MDG NS with MPA1 MEOS for different fixed $\lambda_\Phi \in (2.4\,km, 10^4\,km)$
That for GR is indicated by the black dashed curve.}
\label{Fig4}
\end{minipage}
\end{figure}

The key step in the calculations is to obtain the relation $F_\Lambda(\varphi_c,\xi_c;\lambda_\Phi)=0$ shown in Fig.~\ref{Fig2} for different values
of the Compton length $\lambda_\Phi$.
The boundary conditions at the center of the star
defined by these curves yield the MDG $m_{total} - r^*$ relations shown in Fig.~\ref{Fig3}.
The dashed black line describes the corresponding GR $m^* - r^*$ relation.
As it should be, in the limit $\lambda_\Phi \to 0$
the MDG $m_{total} - r^*$ curves tend to the GR one.
Fig.~\ref{Fig4} shows the corresponding dependencies of the compactness $C=m^*/r^*$ of an MDG NS on the central matter density.
As seen in Figs.~\ref{Fig2}--\ref{Fig4}, the influence of the dark scalar on the interior structure of the NS is significant.
The decrease of the value of $\lambda_\Phi$ leads to a narrowing of the domains
of the corresponding variables. When $\lambda_\Phi$ approaches the critical value $\lambda_\Phi^{crit}$
the curves shrink to a point.
This is an indication of the existence of a bifurcation point of the physical part of the phase space of the system.
For an NS with MPA1 MEOS we obtained numerically $\lambda_\Phi^{crit}\approx 2.1$\ km.

\begin{figure*}[!ht]
\centering
\begin{minipage}{13.cm}
\vskip .truecm
\hskip -7.truecm
\includegraphics[width=6truecm]{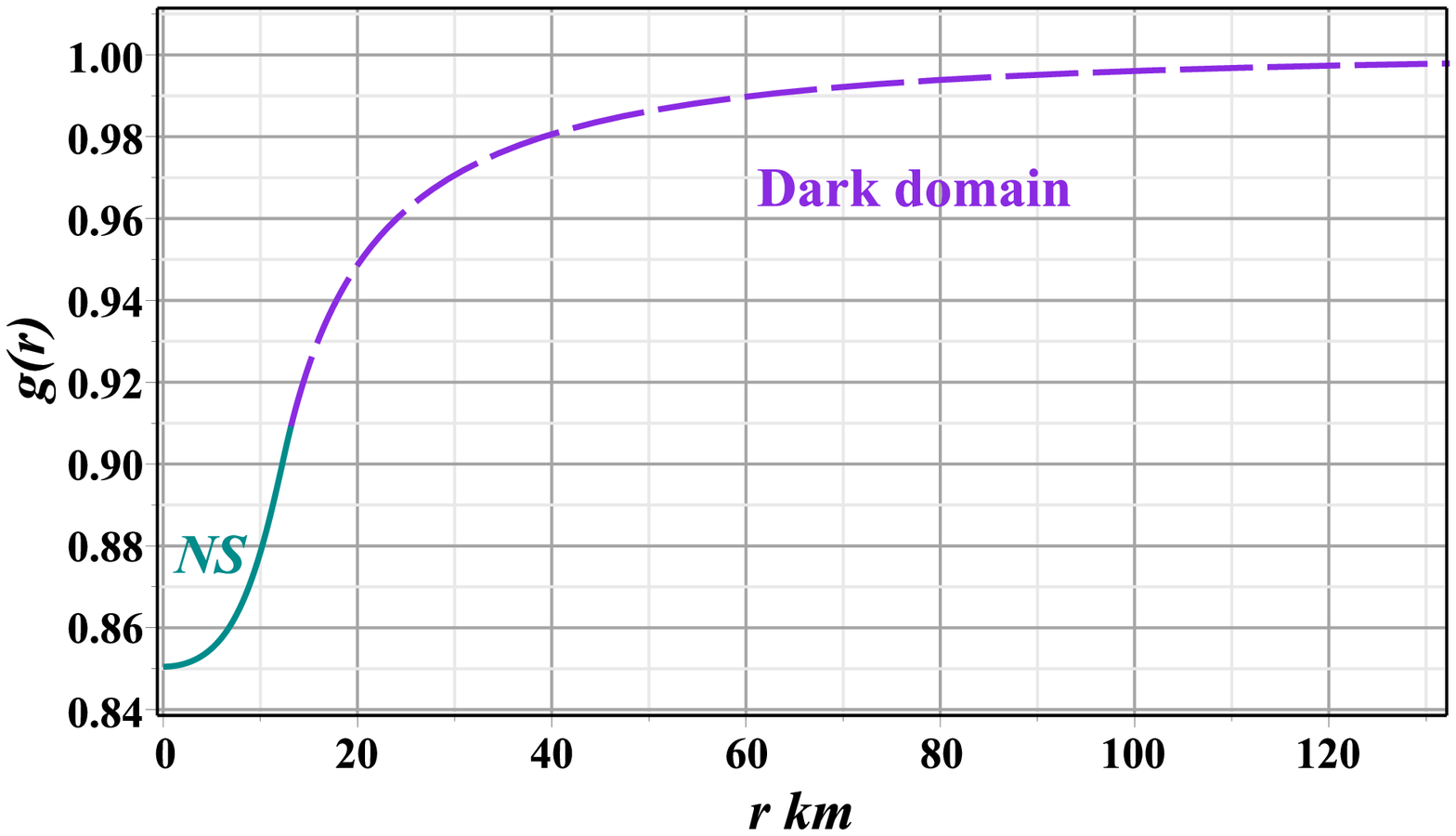}
\end{minipage}
\begin{minipage}{13.cm}
\vskip -3.6truecm
\hskip 7.truecm
\includegraphics[width=6truecm]{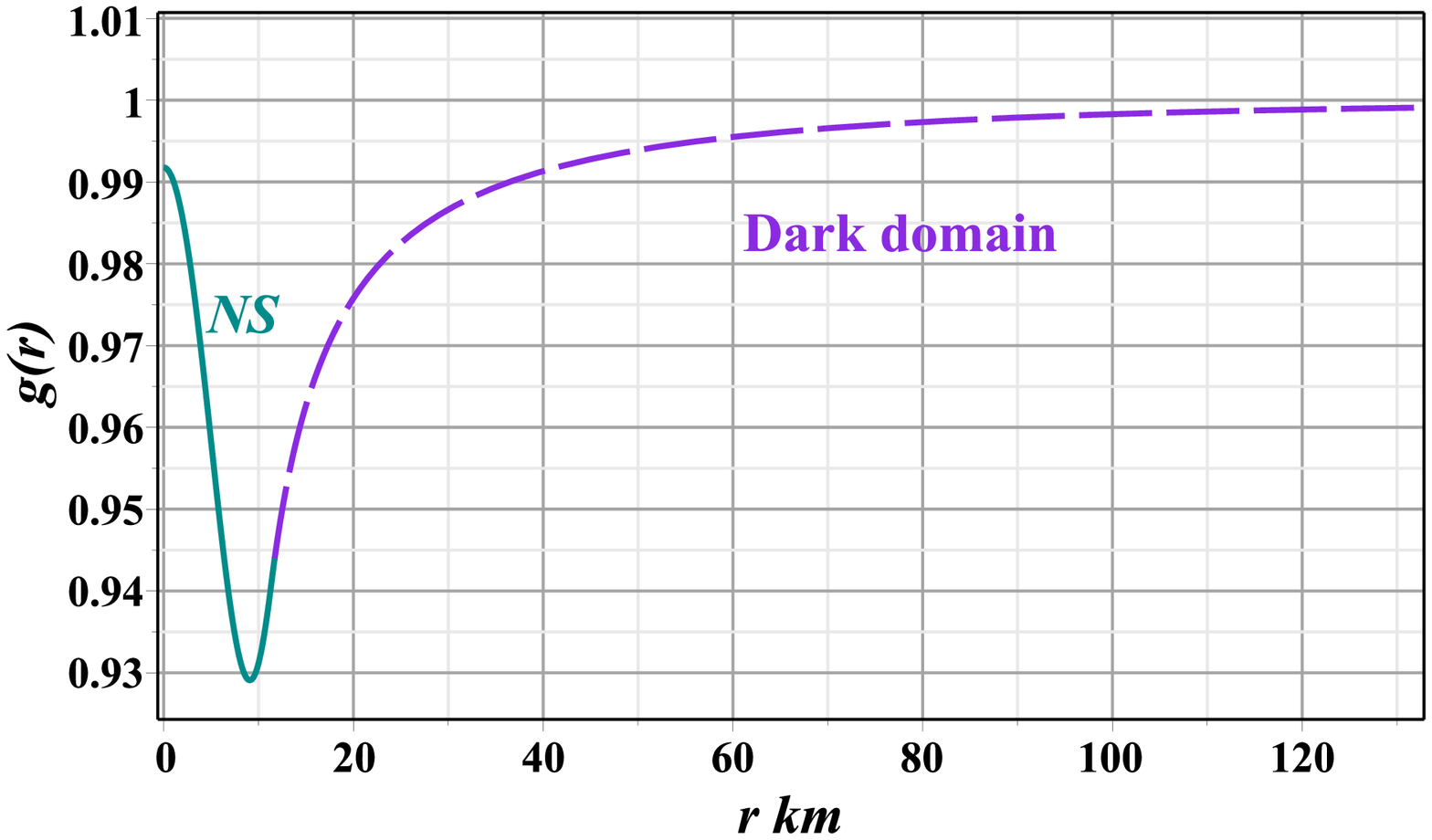}
\end{minipage}
\vskip -.2truecm
\caption{\small Examples of dependence of gravity intensity on $r$.
The dark scalar, having no direct interaction with the matter of SPM,
influences the structure of the NS by changing significantly the intensity of gravity inside it, and in its vicinity.}
\label{Fig5}
\end{figure*}

\begin{figure}[!ht]
\centering
\begin{minipage}{26.cm}
\vskip -.truecm
\hskip -13.truecm
\includegraphics[width=7.5truecm]{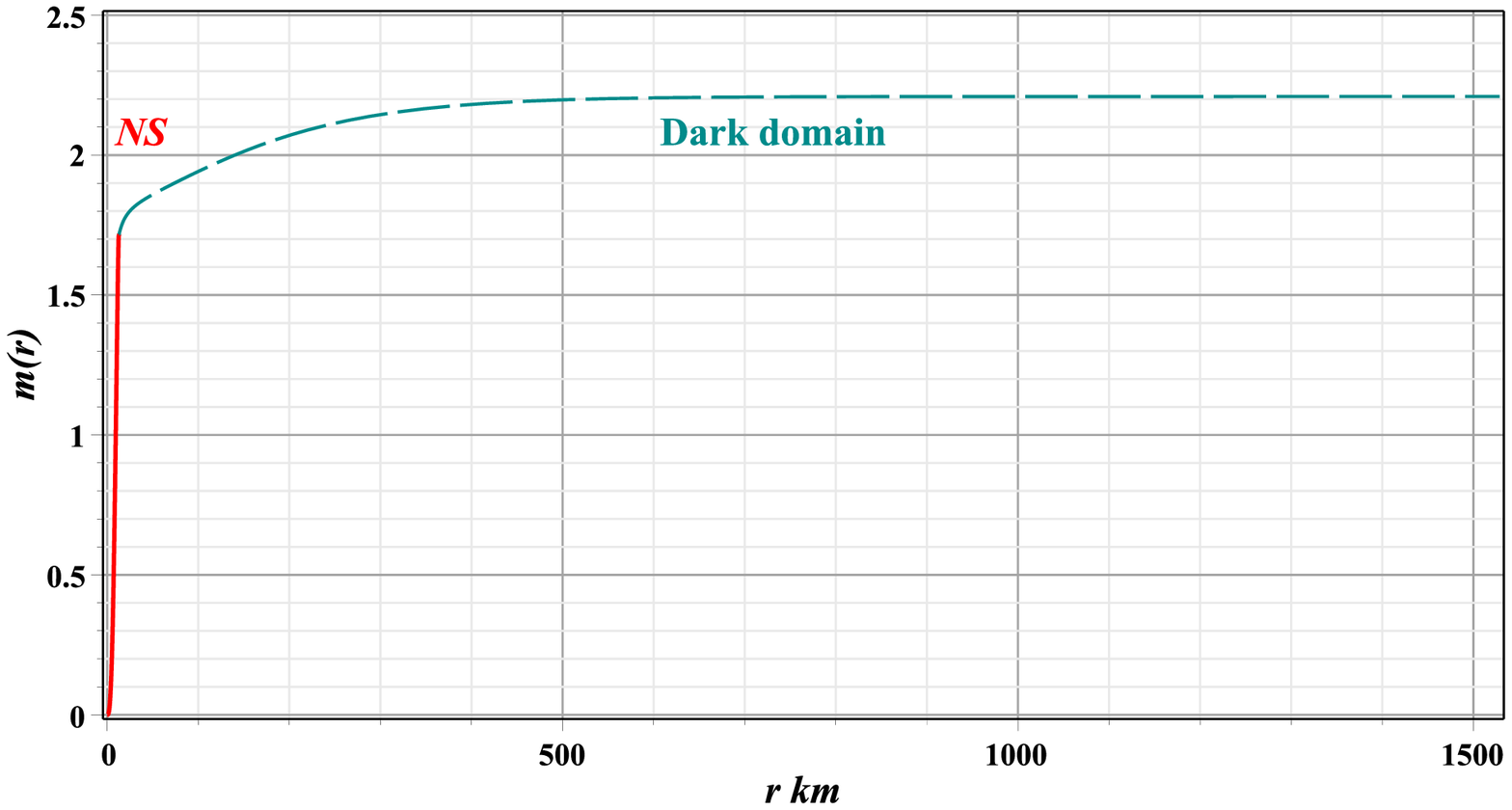}
\end{minipage}
\begin{minipage}{26.cm}
\vskip -3.5truecm
\hskip -15.3truecm
\includegraphics[width=3.truecm]{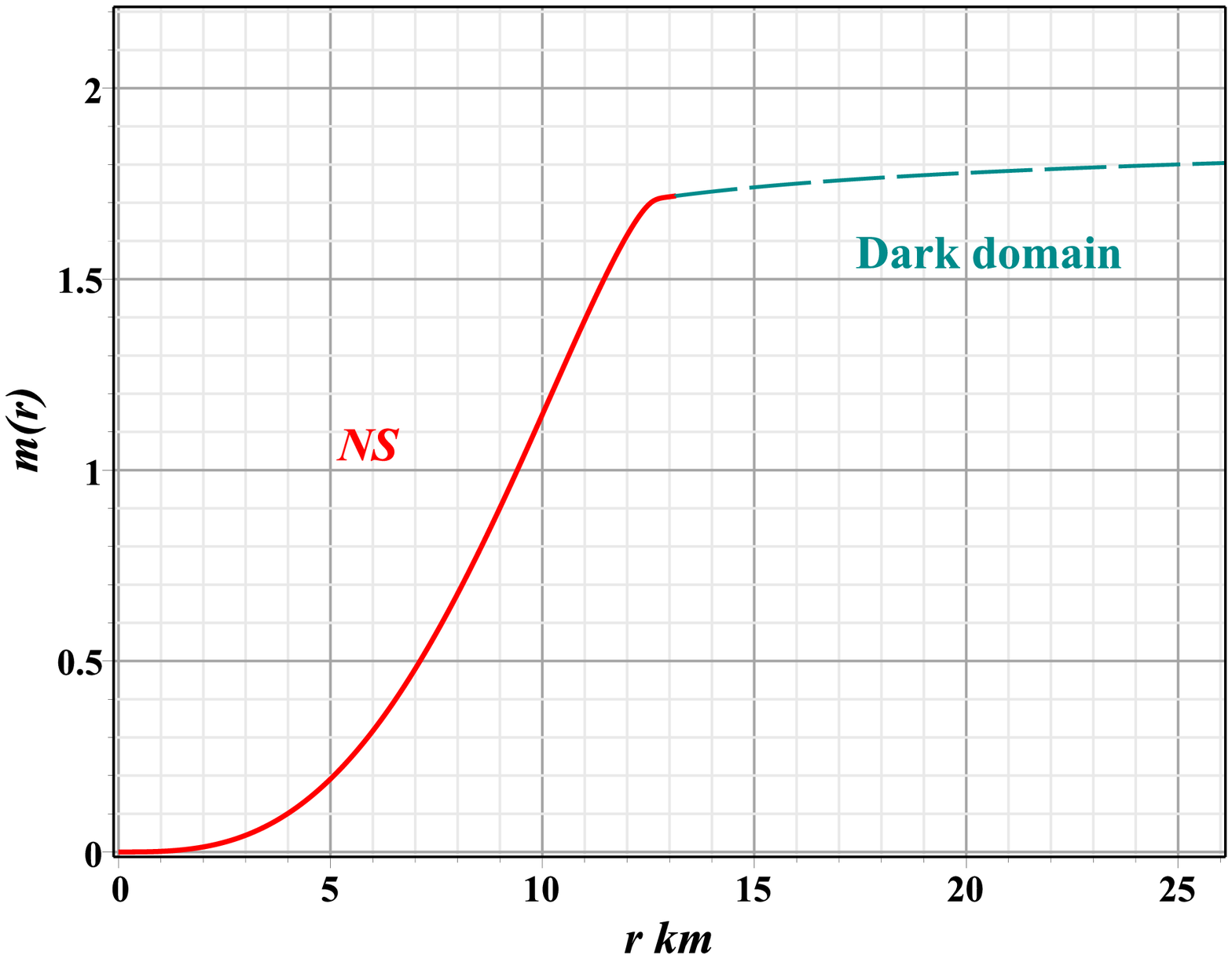}
\end{minipage}
\hskip 20.truecm
\begin{minipage}{26.cm}
\vskip -4.6truecm
\hskip 22.5truecm
\includegraphics[width=4.5truecm]{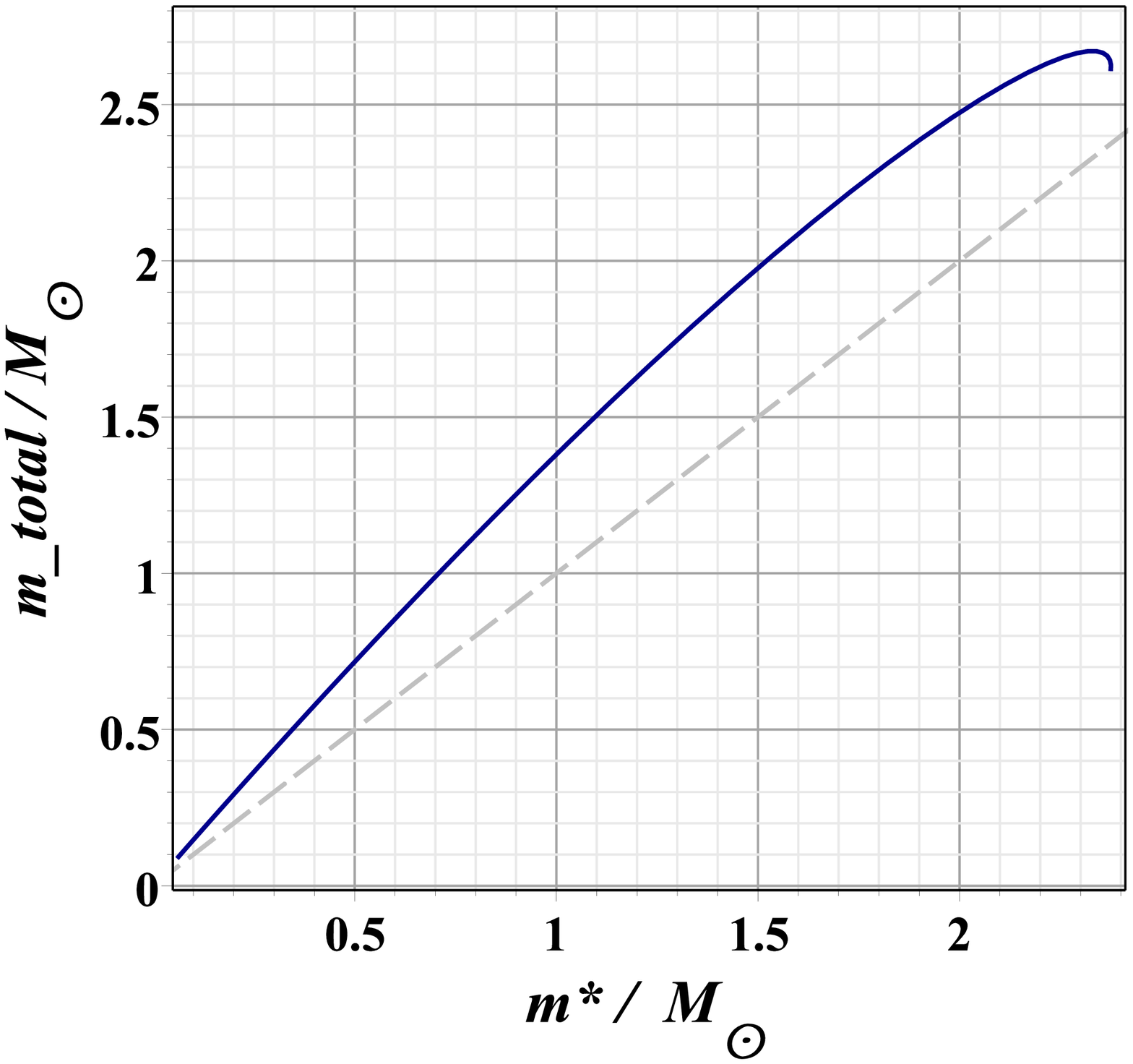}
\end{minipage}
\vskip .truecm
\caption{\small Left: The dependence of the surrounding mass $m(r)$ on the luminosity radius $r$.
                Right: $m_{total}$ versus $m^*$}
\label{Fig6}
\end{figure}

In Figs.~\ref{Fig5}--\ref{Fig6} one can see more consequences of the presence of the dark scalar in NS physics.
%

\begin{figure}[!ht]
\centering
\begin{minipage}{12.cm}
\vskip -.truecm
\hskip -.truecm
\includegraphics[width=7.truecm]{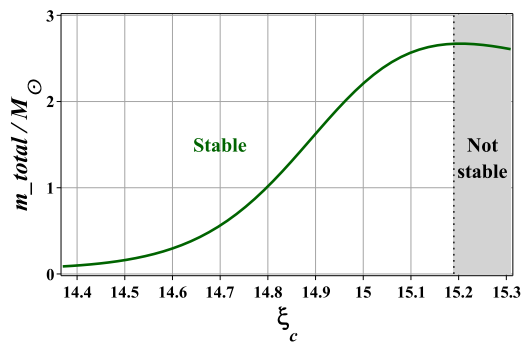}
\vskip -.3truecm
\caption{\small The $m_{total}$ - $\xi_c$ dependence of MDG NS with MPA1 MEOS}
\label{Fig7}
\end{minipage}
\end{figure}

After all, as seen in Fig.~\ref{Fig7}, the MGD NSs have qualitatively the same stability properties as the GR ones.
Clearly, the numbers depend on the dark scalar mass and may vary to some extent.

\section{Discussion}

The present paper is devoted to MDG as the simplest possible solution of the current problems with DM and DE.
Here we considered for the first time a realistic  MDG model of a neutron star (NS) with MPA1 MEOS reaching the following basic results:

The derivation of the MDG generalizations of the TOV equations and the corresponding
boundary conditions at the star's center, at its edge, and at the cosmological horizon of the de Sitter like
MDG Universe.
A clear physical understanding of the SSSS structure is reached using notions of
cosmological energy density and pressure, dilatonic energy density and pressure, and the corresponding equations of state:
CEOS, DEOS, and MEOS.

The maximal mass $M_{max}={ \frac {8\pi} {\kappa \sqrt{\Lambda}} } \sim 10^{22} M\odot$ of any matter object
in an MDG Universe is consistent with observations.

The existence of three very different MDG scales, the Compton length of the scalar
field $\lambda_\Phi \in (10^{-27}, 10^{25})$\ cm, the star's radius $r^* \sim 10^6$\ cm, and
the finite radius of the MDG Universe $r_{U} \sim 10^{28}$\ cm, is a source of numerical difficulties.
Owing to the introduction of a new dark scalar field $\varphi=\ln(1+\ln\Phi)$
we were able to numerically study for the first time
an unprecedentedly large (four orders of magnitude) interval of $\lambda_\Phi$ for neutron stars with MPA1 MEOS
and to discover the existence of $\lambda_\Phi^{crit}\approx 2.1$\ km,
related to the bifurcation of the physical domain in the phase space of the system.
This value corresponds to a critical mass $m_\Phi^{crit} \approx 5\times 10^{-11}$\ eV/$c^2$ and depends on the MEOS of the NS.

The kind of narrowing, typical for a bifurcation point, of the domains of
the corresponding variables (see Figs.~\ref{Fig2}--\ref{Fig4})
may explain the astrophysical observations which do not show the existence of NSs
along the whole theoretical mass--radius curves,
but only on a narrow part of them \cite{Lattimer11,Valentim11,Guillot13,Kiziltan13,Lattimer13}.
This unexpected result of ours may serve as a new criterion for the choice of a realistic MEOS,
coherently and simultaneously with the determination of the dark scalar mass $m_\Phi$.
For this purpose, one needs a more profound application of the mathematical theory of bifurcations
and we intend to publish the corresponding results separately.

The gravitational force in the interior of an MDG star is smaller than in GR and may vary in different ways, see Fig.~\ref{Fig5}.
This result refutes the wide-spread opinion that the $\Lambda$ term in the gravitational action is
inessential at the stellar scale.
Indeed, in the physical de Sitter vacuum, we have  $\Phi=1$ and $U=1$ and
the extremely small observed value of $\Lambda$ makes negligible the $\Lambda$-terms
at the scale of the solar system and in a large enough vicinity outside a star.
However, inside the star, we have no vacuum state and the dilaton deviates from its vacuum value.
As a result, the cosmological potential changes its value very significantly inside the star
and compensates for the small value of the cosmological constant.

The above statements and the previous work on MDG opens a novel possibility:
To look simultaneously and coherently for a realistic MEOS of different physical objects
and for a realistic withholding cosmological potential,
which together are able to describe the variety of phenomena at very different physical scales.

It would be very interesting to work out models of moving and rotating stars in MDG.
In this case, one can expect not only an asymmetric stellar configuration and dark domains,
but also the appearance of different centers of the star and its dark domain,
or even detachment of parts of the dark domain.
At the galactic and galactic cluster scales, such phenomena have already been observed \cite{Clowe06,Graham15,Harvey15,Massey15}
and may allow an MDG explanation.

\acknowledgments
The author is deeply indebted to the Directorate of the
Laboratory of Theoretical Physics, JINR, Dubna, for
good working conditions and support.
He also is grateful to
Alexei Starobinsky, Salvatore Capozziello, Alexander Zacharov,
Luciano Rezzolla, Valeria Ferrari, and Fiorella Burgio for stimulating discussions
on different topics of the present paper.
Special thanks to Kazim Yavuz Ek\c{s}i for providing numerical EoS data.

This research was supported in part by the Foundation for
Theoretical and Computational Physics and Astrophysics
and by the Bulgarian Nuclear Regulatory Agency, Grant for 2015,
as well as by ``NewCompStar,'' COST Action MP1304.

\appendix

\section{The field equations for static spherically symmetric MDG}\label{basicEqsa}
Taking into account the expression for the scalar spacetime curvature
$R\!=\!-e^{-\lambda}\left(\nu^{\prime\prime}+\nu^\prime {\frac {\nu^\prime-\lambda^\prime } 2} +2{\frac {\nu^\prime-\lambda^\prime} r}+{\frac 2 {r^2}}\right)
+{\frac 2 {r^2}}$,
from the basic Eqs. \eqref{DGE} one obtains
the system of ordinary differential equations
\begin{widetext}
\begin{subequations}\label{ODE:abcd}
\ben
\Phi^{\prime\prime}+\left({\frac 2 r}-{\frac 1 2}\left( 3\nu^\prime+\lambda^\prime \right)\right)\Phi^\prime +
\left({\frac 2 {r^2}}-\nu^{\prime\prime} -{\frac 1 2}\nu^\prime\left( \nu^\prime-\lambda^\prime \right)
-{\frac 2 r}\left( \nu^\prime+\lambda^\prime \right)\right)\Phi+\left(24\pi \left(\epsilon+p\right)-{\frac 2 {r^2}}\Phi\right)e^\lambda&=&0,\hskip 1.truecm \label{ODE:a}\\
-3\Phi^{\prime\prime} +\left({\frac 2 r}+{\frac 1 2}\left( \nu^\prime+3\lambda^\prime \right)\right)\Phi^\prime +
\left({\frac 2 {r^2}}-\nu^{\prime\prime} -{\frac 1 2}\nu^\prime\left( \nu^\prime-\lambda^\prime \right)
+{\frac 2 r}\left( \nu^\prime+\lambda^\prime \right)\right)\Phi-\left(8\pi \left(\epsilon+p\right)+{\frac 2 {r^2}}\Phi\right)e^\lambda&=&0, \label{ODE:b}\\
\Phi^{\prime\prime} -\left({\frac 2 r}-{\frac 1 2}\left( \nu^\prime-\lambda^\prime \right)\right)\Phi^\prime -
\left({\frac 2 {r^2}}-\nu^{\prime\prime} -{\frac 1 2}\nu^\prime\left( \nu^\prime-\lambda^\prime \right)
\right)\Phi-\left(8\pi \left(\epsilon+p\right)-{\frac 2 {r^2}}\Phi\right)e^\lambda&=&0, \label{ODE:c}\\
\Phi^{\prime\prime}+\left({\frac 2 r}+{\frac 1 2}\left( \nu^\prime-\lambda^\prime \right)\right)\Phi^\prime +
\left(8\pi \left({\frac \epsilon 3}-p\right)-V_{,\Phi}\right)e^\lambda&=&0. \label{ODE:d}
\een
\end{subequations}
\end{widetext}
The first three of Eqs. \eqref{ODE:abcd} are $\tbinom {t}{t}$, $\tbinom {r}{r}$, and $\tbinom {\theta}{\theta}$  projections of
the first of equations in Eqs. \eqref{DGE},
and Eq. \eqref{ODE:d} follows from the second equation of \eqref{DGE}.

Because the first equation of Eqs. \eqref{DGE} is traceless, Eqs. \eqref{ODE:a}--\eqref{ODE:c}
are not independent and one can omit Eq. \eqref{ODE:c}.
Then, using the relation $R=2 U_{,\Phi}$ \cite{Fiziev13},
the explicit form of the scalar curvature $R$, and the definition of $V_{,\Phi}$, one obtains from
Eqs. \eqref{ODE:a} and \eqref{ODE:b}
\begin{widetext}
\begin{subequations}\label{ODE2:ab}
\ben
\Phi^{\prime\prime}-
\left({\frac {\lambda^\prime} 2} -{\frac 2 r}\right)\Phi^\prime -\left({\frac {\lambda^\prime} r} -{\frac 1 {r^2}}\right)\Phi
+\left(U-{\frac \Phi {r^2}}+8\pi \epsilon\right)e^\lambda&=&0, \label{ODE2:a}\\
\left({\frac {\nu^\prime} 2} +{\frac 2 r}\right)\Phi^\prime +\left({\frac {\nu^\prime} r} +{\frac 1 {r^2}}\right)\Phi+\left(U-{\frac \Phi {r^2}}-8\pi p\right)e^\lambda&=&0.
\label{ODE2:b}
\een
\end{subequations}
\end{widetext}

Let us introduce the mass function $m(r)$ obeying the relations
$e^\lambda=\Delta /r$, $\Delta(r)=r-2 m(r)-{\frac \Lambda 3}r^3$.
The next step is to define the dilaton pressure and dilaton energy density according to relations \eqref{epP}.
As a result, Eq. \eqref{ODE2:a} acquires the form \eqref{DE:a}.
Using Eqs. \eqref{epP} we obtain also the second equation of \eqref{DE:b}.

Using the definitions of $p_{eff}$ and $p_\Lambda$,
from Eq. \eqref{ODE2:b} we obtain an equation
for the function $\nu(r)$ that is valid both inside and outside SSSS:
\ben
\nu^\prime=  {\frac 2 r}{\frac { m+4\pi r^3 p_{eff}/\Phi }{\Delta -2 \pi r^3 p_\Phi/\Phi}}.
\la{dnu}
\een

Outside the star, the function $\nu(r)$ is determined by Eq.~\eqref{dnu} under the additional conditions $p \equiv 0$, $\epsilon \equiv 0$.

Inside the star we use the standard approach:
in MDG, as well as in GR, the equation of motion of matter is $\nabla_\nu T_\mu^\nu=0$.
We assume that SSSS is filled by a standard ideal fluid with energy--momentum tensor
$T_\mu^\nu=(\epsilon+p)u_\mu u^\nu - p\delta_\mu^\nu$ at rest,
i.e., when $u_\mu=\delta_\mu^0$.
Then, in thermodynamical equilibrium, inside the star we have the usual relation
$p^\prime = - (\epsilon+p)\nu^\prime/2$.
Inserting Eq. \eqref{dnu} into it,
we obtain the second of Eqs. \eqref{DE:a},
thus arriving at the extension \eqref{DE:ab} of the  TOV equations in the case of MDG SSSS.

Finally, in terms of these newly introduced quantities, Eq. \eqref{ODE:d} transforms into DEOS.

\vskip 1.truecm

\section{Conditions at the center of the star and the behaviour of the solutions}\label{basicEqsb}

As in Newtonian gravity, and in the GR gravity of SSSS,
we assume that all physical quantities are finite at the center $r_c=0$ of the MDG star.
The withholding property of the cosmological potential $U(\Phi)$ dynamically ensures that $0 < \Phi_c < \infty$
\cite{Fiziev13} and, as a result of their definitions,
$U_c$, $C_c$, $V_{,\Phi}^c$ , $\epsilon_\Lambda^c$, and $p_\Lambda^c$  are finite.
As usual, we assume $0<\epsilon_c<\infty$ and $0<p_c<\infty$ for the stellar matter.
In addition, we require $|p_\Phi^c|<\infty$ and $|\epsilon_\Phi^c|<\infty$.
Then $|p_{eff}^c|<\infty$ and $|\epsilon_{eff}^c|<\infty$.

Now, taking the limit $r\to 0$, we obtain from the first of Eqs. \eqref{DE:a}
$m^\prime\sim {\tfrac {4\pi r^2}{\Phi_c}}\epsilon_{eff}^c,\quad \Rightarrow
m(r)\sim m_c+{\tfrac {4}{3}}\pi r^3\epsilon_{eff}^c/\Phi_c$.
If the integration constant $m_c\neq 0$, then the second of  Eqs. \eqref{DE:a}
yields
$p(r)\sim {\frac 1 2}\left( p_c+\epsilon_c\right)\ln(r/r_0)+\text{const}$.
Since $p_c+\epsilon_c >0$, then $p(r) \to -\infty$ for $r \to 0$.
This is physically unacceptable, just as in Newtonian gravity and in GR.
Thus we see that one must suppose that $m_c \equiv 0$. Then
\ben
m(r) \sim {\tfrac {4}{3}} \pi r^3 \epsilon_{eff}^c /\Phi_c +{\cal{O}}(r^4),\quad
p(r) \sim p_c-{\tfrac {2\pi}{\Phi_c}} \left(p_c+\epsilon_c\right)
\left(p_{eff}^c +{\tfrac 1 3}\epsilon_{eff}^c\right)r^2+{\cal{O}}(r^3).
\la{m_p_to_0}
\een
From the second of Eqs. \eqref{DE:b} we obtain
$p_\Phi^\prime \sim -{\tfrac 3 r}p_\Phi^c-{\tfrac 2 r}\epsilon_\Phi^c,\quad \Rightarrow
p_\Phi(r)\sim -\left(3p_\Phi^c+2\epsilon_\Phi^c\right)\ln(r/r_0)+\text{const}$.
Obviously, the only physical solution with $|p_\Phi^c| <\infty$ is the one with
\ben
p_\Phi^c = -{\tfrac{2}{3}}\epsilon_\Phi^c=
{\frac 2 9}T_c -{\frac{\Lambda}{12\pi}}V_{\!{}_{,\Phi}}(\Phi_c),\quad T_c=\epsilon_c- 3p_c.
\la{pP_to_0}
\een

Using this result, from Eqs. \eqref{DE:b} we obtain
\ben
\Phi(r) \sim \Phi^c+{\tfrac{4}{3}}\pi r^2\epsilon_\Phi^c+{\cal{O}}(r^3)=
\Phi_c\left( 1+{\tfrac{\epsilon_\Phi^c}{\epsilon_{eff}^c}}\,{\frac{m(r)} r}\right)+{\cal{O}}(r^3),
\quad p_\Phi(r)\sim p_\Phi^c+{\cal{O}}(r^2).
\la{Phi_to_0}
\een



\begin{thebibliography}{28}

\bibitem{Planck_XIV_15} Planck collaboration, {\em Planck 2015 results. XIV. Dark energy and modified gravity}, arXiv:1502.01590 (2015).

\bibitem{Buhdahl72} H. A. Buhdahl, MNRAS {\bf 150} 1 (1970).

\bibitem{Starobinsky80} A. A. Starobinsky, Phys. Lett. B {\bf 91}  99 (1980).

\bibitem{Muiller88} V. M\"uller, H.-J. Schmidt, A. A. Starobinsky, Phys. Lett. B {\bf 202} 198 (1988).

\bibitem{Nojiri03} S. Nojiri, Sergei D. Odintsov, Phys. Rev. D {\bf 68} 123512 (2003).

\bibitem{Carroll04} S. M. Carroll, V. Duvvuri, M. Trodden, M. S. Turner, Phys. Rev. D {\bf 70} 043528 (2004).

\bibitem{Nojiri04} S. Nojiri, S. D. Odintsov, Gen. Rel. Grav. {\bf 36} 1765 (2004).

\bibitem{Abdalla05} M. C. B. Abdalla, S. Nojiri, S. D. Odintsov, Class. Quant. Grav. {\bf 22} L 35 (2005).

\bibitem{Capozzielo06a} S. Capozziello, S. Nojiri, S. D. Odintsov, Phys. Lett. B {\bf 634} 93 (2006).

\bibitem{Capozziello06b} S. Capozziello, S. Nojiri, S. D.  Odintsov, A. Troisi Phys. Lett. B {\bf 639} 135 (2006).

\bibitem{Faraoni06} V. Faraoni, Phys. Rev. D {\bf 74} 104017 (2006).

\bibitem{Starobinsky07} A. A. Starobinsky, JETP Lett. {\bf 86} 157 (2007).

\bibitem{Amendola07} L. Amendola, R. Gannouji, D. Polarski, S. Tsujikawa, Phys. Rev. D {\bf 75} 083504 (2007).

\bibitem{Appleby07} S. A. Appleby, R. A. Battye, Phys. Lett. B {\bf  654} 7 (2007).

\bibitem{Hu07} W. Hu, I. Sawicki, Phys. Rev. D  {\bf 76}   064004 (2007).

\bibitem{Li07} B. Li, J. D. Barrow, Phys. Rev. D {\bf 75} 084010 (2007).

\bibitem{Bamba08} K. Bamba, S. Nojiri, S. D. Odintsov, JCAP {\bf 0810} 045 (2008).

\bibitem{Nojiri08} S. Nojiri, S. D. Odintsov, lectures given at JGRG17 (Nagoya, Japan) and at VI Winter School on Theor. Phys. (Dubna, Russia), arXiv:0801.4843

\bibitem{Amendola08} L. Amendola and S. Tsujikawa, Phys. Lett. B {\bf 660} 125 (2008).

\bibitem{Tsujikawa08} S. Tsujikawa, Phys. Rev. D {\bf 77} 023507 (2008).

\bibitem{Nojiri08c} S. Nojiri, S. D. Odintsov, Phys. Rev. D {\bf 77} 026007 (2008).

\bibitem{Cognola08} G. Cognola, E. Elizalde, S. Nojiri, S. D.  Odintsov, L. Sebastiani, S. Zerbini, Phys. Rev. D{\bf 77} 046009 (2008).

\bibitem{Starobinsky10} S. Appleby, R. Battye, A. Starobinsky, JCAP {\bf 1006} 005 (2010); arXiv:0909.1737.

\bibitem{Felice10} A. De Felice, S. Tsujikawa, Living Rev. Rel. {\bf 13} 3 (2010).

\bibitem{Sotiriou10} T. P. Sotiriou, V. Faraoni, Rev. Mod. Phys. {\bf 82} 451 (2010).

\bibitem{Tsujikawa10} S. Tsujikawa, Lect. Notes Phys. {\bf 800} 99 (2010).

\bibitem{Motohashi10} H. Motohashi, A. A. Starobinsky, J. Yokoyama, Progr. Theor. Phys. {\bf 124} 541 (2010).

\bibitem{Capozziello11} S. Capozziello, M. De Laurentis, Physics Reports {\bf 509}, 167 (2011).

\bibitem{Nojiri11} S. Nojiri, S. D. Odintsov, Physics Reports {\bf 505} 59 (2011).

\bibitem{Capozziello11a} S. Capozziello, V. Faraoni, {\it Beyond Einstein Gravity},
Fundamental Theories of Physics {\bf 170}, Springer, 2011.

\bibitem{Rubakov11} D. S. Gorbunov, V. A. Rubakov {\em Introduction to the Theory of the Early Universe: Hot Big Bang Theory}, World Scientific, (2011).

\bibitem{Clifton12} T. Clifton, P. G. Ferreira, A. Padill, C. Skordis, Physics Reports {\bf 513} 1 (2012).

\bibitem{Bamba12} K. Bamba, S. Capozziello, S. Nojiri, S. D. Odintsov, Astrophysics and Space Science  {\bf 342} 155 (2012).

\bibitem{Starobinsky15} A. S. Chudaykin, D. S. Gorbunov, A. A. Starobinsky, R. A. Burenin, JCAP {\bf 05} 004 (2015); arXiv:1412.5239.


\bibitem{Gannouji12} R. Gannouji, M. Sami, I. Thongkool, Phys. Lett. B {\bf 716} 255 (2012); arXiv:1206.3395.


\bibitem{OHanlon72} O'Hanlon, Phys. Rev. Lett. {\bf 29} 137 (1972).

\bibitem{Fiziev00a} P. Fiziev, Mod. Phys. Lett. A {\bf 15} 1077 (2000).

\bibitem{EFarese01} G. Esposito-Farese, D. Polarski, Phys. Rev. D {\bf 63} 063504  (2001).

\bibitem{Fiziev02} P. Fiziev,  arXiv:gr-qc/0202074.

\bibitem{Fiziev03} P. Fiziev, D. Georgieva, Phys. Rev. D {\bf 67} 064016 (2003).

\bibitem{Fujii03} Y. Fujii, K. Maeda, {\em The Scalar-Tensor Theory of Gravitation}, Cambridge 2003.

\bibitem{Fiziev13} P. Fiziev, Phys. Rev. D {\bf 87} 0044053 (2013).


\bibitem{Berti2015} E. Berti et al., Class. Quant. Grav. (2015); arXiv:1501.07274.

\bibitem{Burgio2015} F. Burgio, ``The Equation of State of Neutron Star Matter,'' talk at the Conference
Annual NewCompStarConference, 15--19 June 2015, Budapest, http://indico.kfki.hu/event/254/session/9/contribution/155/material/slides/0.pdf

\bibitem{Fiziev14a} P. Fiziev, arXiv:1402.2813.

\bibitem{Fiziev14b} P. Fiziev, 	Symposium ``Frontiers of Fundamental Physics 14,'' Marseille, France, July 15--18, 2014, PoS(FFP14) 080; arXiv:1411.0242;


\bibitem{Fiziev15a} P. Fiziev, K. Marinov, Bulgarian Astronomical Journal {\bf 23} 3, (2015); arXiv:1412.3015.

\bibitem{Fiziev15b}  P. Fiziev, {\em A realistic model of a neutron star in a modified theory of gravity
},  talk at the Annual NewCompStar Conference, 15--19 June 2015, Budapest,
http://indico.kfki.hu/event/254/session/5/contribution/134/material/slides/0.pdf




\bibitem{Planck_XX_15} Planck collaboration,  {\em Planck 2015 results. XX. Constraints on inflation},arXiv:1502.02114 (2015).

\bibitem{Fujii71} Y. Fujii, Nature, Physical Science {\bf 234} 5 (1971).

\bibitem{Murata15} J. Murata, S. Tanaka, Class. Quantum Grav.  {\bf 32} 033001, 2015; arXiv:1408.3588.



\bibitem{Kainulainen07a} K. Kainulainen, J. Piilonen, V. Reijonen, D. Sunhede, Phys. Rev. D {\bf 76} 024020 (2007).

\bibitem{Kainulainen07b} K. Kainulainen, V. Reijonen, D. Sunhede, Phys. Rev. D {\bf 76} 043503 (2007).

\bibitem{Multamaki07} T. Multamaki, I. Vilja, Phys. Rev. D, {\bf 76} 064021 (2007).

\bibitem{Nojiri07} S. Nojiri, S. D. Odintsov, Int. J. Geom. Meth. Mod. Phys. {\bf 4} 115 (2007).


\bibitem{Multamaki08} 	T. Multamaki, I. Vilja, Phys. Lett. B {\bf 659} 843 (2008).

\bibitem{Frolov08} A. V. Frolov, Phys. Rev. Lett. {\bf 101} 061103 (2008).

\bibitem{Kobayashi08} T. Kobayashi, K. Maeda, Phys. Rev. D {\bf 78} 064019 (2008).

\bibitem{Henttunen08} K. Henttunen, T. Multamaki, I. Vilja,  Phys. Rev. D {\bf 77} 024040 (2008).

\bibitem{Kainulainen08} K. Kainulainen, D. Sunhede, Phys. Rev. D {\bf 78} 063511 (2008).


\bibitem{Nojiri09} S. Nojiri, S. D. Odintsov, AIP Conf. Proc. {\bf 1115} 212 (2009).

\bibitem{Babichev09} E. Babichev, D. Langlois, Phys. Rev. D {\bf 80}, 121501 (2009).



\bibitem{Babichev10} E. Babichev, D. Langlois, Phys. Rev. D {\bf 81}, 121051 (2010).

\bibitem{Cooney10} A. Cooney, S. DeDeo, D. Psaltis, Phys. Rev. D {\bf 82}, 064033 (2010).


\bibitem{Capozziello11b} S. Capozziello, M. De Laurentis, S. D. Odintsov, A. Stabile, Phys. Rev. D {\bf 83}, 064004 (2011).

\bibitem{Arapoglu11} A. S. Arapoglu, C. Deliduman, K. Y. Eksi,  J. Cosmology and Astroparticle Physics {\bf 07} 020 (2011).


\bibitem{Santos12} E. Santos, Astrophys. Space Sci. {\bf 341} 411 (2012).

\bibitem{Ang12} Ang Li, Feng Huang, Ren-Xin Xu, Astroparticle Phys. {\bf 37} 70 (2012).

\bibitem{Deliduman12} C. Deliduman, K. Y. Eksi, V. Kekes, JCAP {\bf 5} 036 (1012).


\bibitem{Alavirad13} H. Alavirad, J. M. Weller,  Phys. Rev. D {\bf 88}, 124034 (2013).

\bibitem{Astashenok13} A. V. Astashenok, S. Capozziello, S. D. Odintsov,  arXiv:1309.1978.

\bibitem{Ganguly13} A. Ganguly, R. Gannouji, R. Goswami, S. Ray, arXiv:1309.3279.

\bibitem{Orellana13} M. Orellana, F. G. Florencia, A. T. Pannia, G. E. Romero, Gen. Rel. Grav. {\bf 45} 771 (2013).



\bibitem{Astashenok14} A. V. Astashenok, S. Capozziello, S. D. Odintsov, Phys. Rev. D {\bf 89} 103509 (2014); arXiv:1401.4546.


\bibitem{Astashenok15} A. V., Astashenok, S. Capozziello, S. D. Odintsov, Phys.Lett. B {\bf 742} 160 (2015);arXiv:1412.5453.

\bibitem{Wetterich15} C. Wetterich,  Lect. Notes in Physics {\bf 892} 57 (2015). Springer; arXiv:1402.5031.

\bibitem{Khoury04} J. Khoury, A. Weltman, Phys. Rev. Lett. {\bf 93}, 171104 (2004);
J. Khoury, A. Weltman, Phys. Rev. D {\bf 69}, 044026 (2004).

\bibitem{Brax15} Philippe Brax, Anne-Christine Davis, arXiv:1506.01519.

\bibitem{Holz12}D. E. Holz, S. Perlmutter, The Astrophysical Journal Letters {\bf 755} L36, (2012).

\bibitem{MPA1a} H. M\"uther, M. Prakash, T. L. Ainsworth, Phys. Lett. B {\bf 199} 469 (1987).

\bibitem{MPA1b} M. Prakash and T. L. Ainsworth, Phys. Rev. C {\bf 36} 346 (1987).

\bibitem{MPA1c} Can Gungor, K. Yavuz Eksi, Conference proceedings of ``Advances in Computational Astrophysics: Methods, tools and outcomes''
            Cefalu (Sicily, Italy) June 13--17, 2011;	arXiv:1108.2166.

\bibitem{Lattimer11}	James M. Lattimer, M. Prakash, {\em What a Two Solar Mass Neutron Star Really Means}, in From Nuclei to Stars, ed. S. Lee, p. 275. WorldScientific. (2011); arXiv:1012.3208.

 \bibitem{Valentim11}	 R. Valentim, E. Rangel, J. E. Horvath, MNRAS {\bf 414} 1427 (2011); arXiv:1101.4872.

\bibitem{Guillot13} S. Guillot, M. Servillat, N. A. Webb, R. E. Rutledge, ApJ. {\bf 772} 7 (2013);  arXiv:1302.0023;

\bibitem{Kiziltan13} B. Kiziltan, A. Kottas, M. De Yoreo, S. E. Thorsett, ApJ {\bf 778} 66 (2013); arXiv:1309.6635.

\bibitem{Lattimer13} J. M. Lattimer, Annu. Rev. Nucl. Part. Sci. {\bf 62} 485 (2012); arXiv:1305.3510.

\bibitem{Clowe06} D. Clowe, M. Bradac, A. H. Gonzalez, M. Markevitch, S. W. Randall, C. Jones, D. Zaritsky, The Astrophysical Journal {\bf 648} L109 (2006).

\bibitem{Graham15} P. W. Graham, S. Rajendran, K. Van Tilburg, T. D. Wiser, Phys. Rev. D {\bf 91} 104003 (2015);


\bibitem{Harvey15} D. Harvey, R. Massey, T. Kitching, A. Taylor, E. Tittley, Science {\bf 347} 1462 (2015); arXiv:1503.07675.

\bibitem{Massey15} R. Massey et al., MNRAS {\bf 449}  3393 (2015); arXiv:1504.03388.

\end{thebibliography}
\end{document}